\documentclass[usenatbib,useams]{mnras}
\usepackage{times, aas_macros, graphicx, amssymb}
\usepackage{mathrsfs}

\pdfoutput=1
\defcitealias{bekki2013}{B13}
\defcitealias{bekki2014}{B14}

\def \Msun{ {\rm M}_{\sun} }

\def \nB{ n_{\rm B} }
\def \nD{ n_{\rm D} }
\def \SigB{ \Sigma_{\rm B} (R) } 
\def \SigD{ \Sigma_{\rm D} (R) } 
\def \deg{ ^{\circ} }
\def \H2{ {\rm H}_2 }

\def \brf{ }
\def \brg{ }

\title[Formation of S0s from compact ellipticals]{Formation of S0s via disc accretion around high-redshift compact ellipticals}

\author[J. Diaz et al.]
{J.D. Diaz$^1$\thanks{Email: jonathan.diaz@uwa.edu.au},
   Kenji Bekki,$^1$ Duncan A. Forbes,$^2$ Warrick J. Couch,$^{2,3}$
   \newauthor
   Michael J. Drinkwater$^4$ and Simon Deeley$^4$ \\
   ${}^1$ICRAR, M468, The University of Western Australia 35 Stirling Highway, Crawley Western Australia, 6009, Australia \\
   ${}^2$Centre for Astrophysics \& Supercomputing, Swinburne University, Hawthorn, VIC 3122, Australia \\
   ${}^3$Australian Astronomical Observatory, 105 Delhi Rd, North Ryde, NSW 2113, Australia \\
   ${}^4$School of Mathematics and Physics, University of Queensland, QLD 4072, Australia \\ 
}

\date{}

\begin{document}

\maketitle
\begin{abstract}

We present hydrodynamical N-body models which demonstrate that elliptical galaxies can transform into S0s by acquiring a disc. In particular, we show that the merger with a massive gas-rich satellite can lead to the formation of a baryonic disc around an elliptical.  We model the elliptical as a massive, compact galaxy which could be observed as a `red nugget' in the high-$z$ universe.  This scenario contrasts with existing S0 formation scenarios in the literature in two important ways.  First, the progenitor is an elliptical galaxy whereas scenarios in the literature typically assume a spiral progenitor.  Second, the physical conditions underlying our proposed scenario can exist in low-density environments such as the field, in contrast to scenarios in the literature which typically address dense environments like clusters and groups.  As a consequence, S0s in the field may be the most likely candidates to have evolved from elliptical progenitors.  Our scenario also naturally explains recent observations which indicate that field S0s may have older bulges than discs, contrary to cluster S0s which seem to have older discs than bulges.

\end{abstract}

\begin{keywords}
   galaxies: elliptical and lenticular -- galaxies: kinematics and dynamics -- galaxies: dwarf
\end{keywords}

\section{Introduction} \label{sec:int}

Theories of galaxy transformation seek to explain how physical processes can drive the evolution of galaxies from one type to another. Lenticular galaxies (also known as `S0s') are well-suited to studies of galaxy transformation because they exhibit various properties of other galaxy types. Like spiral galaxies, S0s have prominent stellar discs supported by rotation.  However, S0s contain little gas and lack spiral structure.  Like ellipticals, S0s are dominated by their pressure-supported bulges. Ever since \citet{hubble36} first hypothesized S0s as a transition type between early and late type galaxies, theorists and observers have sought to explain S0s as the result of transitions between galaxy types. For instance, modern integral field studies have highlighted S0s as an intermediate kinematic class linking nonrotating early-type galaxies and spiral systems (e.g. \citealt{cappellari2011}).

\citet{dressler80} showed that S0s are an abundant galaxy type in dense cluster environments with a number density that grows at the expense of spiral systems.  Consequently, the formation of S0s has long been thought to be the result of physical processing of spiral progenitors within clusters and massive groups.  In the popular `faded spiral' scenario, a spiral galaxy slowly fades and fails to maintain its spiral structure following the removal of its reservoir of gas (e.g. \citealt{larson80}). { \brf A variety of S0 formation scenarios have been proposed, each invoking spiral progenitors: ram pressure stripping (\citealt{quilis2000}), truncation of star formation plus central gas flows (\citealt{johnston2014}), tidal interaction within dense environments (\citealt{bekkicouch2011, byrd90}), and mergers between spirals and other galaxy types (\citealt{borlaff2014, prieto2013, tapia2017}). }

{ \brf Environment is an important factor in characterizing the formation and evolution of S0s, particularly because many of the aforementioned scenarios invoke dense environments such as clusters and groups.  S0s observed in the field are less abundant than their counterparts in clusters and must be associated with distinct evolutionary paths. One such solution is the merger of two spiral galaxies, which can produce an S0-like merger remnant in the field (e.g. \citealt{que2015, bekki98}).

Moreover, recent observations suggest that field S0s may be structurally distinct from their counterparts in dense environments.  S0s in groups and clusters have bulges that are typically younger than their discs (\citealt{johnston2012}), but the opposite appears to be true for the field: S0s observed by \citet{tabor2017} have bulges which are older or similar in age to their discs.  If this distinction between field and cluster S0s is confirmed by further observational studies, it will provide a key ingredient to understanding the distinct formation histories of S0s in the field.

In this paper we postulate a physical mechanism for S0 formation which is unique in three ways: (1) the main progenitor is not a spiral in contrast with other S0 formation scenarios in the literature, (2) the bulge of the final S0 is older than the disc, and (3) the transformation can occur in the field.  In particular, } we show that a massive compact elliptical galaxy can acquire a disc following the merger of a {\brf gas-rich } satellite, effectively transforming it to an S0. { \brf The compact elliptical evolves passively to become the bulge of the S0, and the destruction of the satellite feeds the buildup of a young disc.}

In our scenario, the elliptical progenitor does not resemble compact ellipticals in the local universe such as M32 but is more akin to the massive, compact `red nuggets' that have been recently discovered in the high-$z$ universe ($z>2$) (e.g., \citealt{damjanov2011}).  Several observational studies have previously suggested that some high-redshift compact ellipticals could have evolved into disc galaxies with compact cores (\citealt{dullograham2013, graham2015, delarosa2016}). In addition, various studies indicate that some cores of present-day systems can be linked to high-redshift compact galaxies, including analyses of cosmological simulations (e.g. \citealt{wellons2016}) as well as observations of nearby galaxies (e.g. \citealt{yildirim2017}). In this context, the present work stands as a concrete illustration of one possible evolutionary path for the transformation of compact ellipticals in low density environments.

The structure of our paper is as follows.  In Section \ref{sec:model}, we describe the details of our numerical method.  In Section \ref{sec:res}, we demonstrate the viability of our new formation scenario by presenting a fiducial model.  We provide a discussion in Section \ref{sec:dis}, and we summarize in Section \ref{sec:sum}. { \brf Several alternate models are presented in Appendix \ref{sec:alt} to consider the sensitivity of the adopted model to different physical conditions.}

\section{Numerical Model} \label{sec:model}

\subsection{A new scenario}

We here consider that S0s can be formed from mergers between compact ellipticals and gas-rich low-mass disc galaxies. At high-$z$, this scenario could correspond to the merger of a red nugget and a gas-rich satellite.  { \brf S0s formed by this scenario never pass through a phase as a spiral galaxy, which contrasts with previous scenarios in the literature in which the main progenitor is a spiral galaxy. }

\subsection{Adopted code for dusty hydrodynamics}

We adopt the simulation code that was originally developed in our previous work (\citealt{bekki2013}, hereafter \citetalias{bekki2013}; \citealt{bekki2014}, hereafter \citetalias{bekki2014}).  Applied to the formation of S0s, this code allows us to derive the structural and  kinematical properties, dust abundances, and spatial distributions of atomic and molecular hydrogen (${\rm H_2}$).  We here describe the code only briefly.  A more comprehensive discussion of the adopted code can be found in \citetalias{bekki2013} and \citetalias{bekki2014}.
The code is designed to run on GPU clusters so that the most time-consuming
part of the simulation (i.e., gravitational calculations) can be accelerated on the specialized hardware of GPUs.  In contrast, the gas dynamics, dust evolution,  conversion from H~{\sc i} to ${\rm H_2}$, and star formation are performed on CPUs. The smoothed-particle hydrodynamics (SPH) method is adopted to follow the time evolution of gas dynamics. 

Using the new code, we can investigate the following physical processes in a self-consistent manner: gas dynamics, star formation, ${\rm H_2}$ formation on dust grains, formation of dust grains in the stellar winds of supernovae (SNe) and asymptotic giant branch (AGB) stars, 
time evolution of interstellar radiation field (ISRF), growth and destruction processes
of dust in the interstellar medium (ISM), and ${\rm H_2}$ photo-dissociation due to far ultra-violet (FUV) light. The code does not include feedback from active galactic nuclei (AGN)
on the ISM nor the growth of supermassive black holes (SMBHs). Since such AGN feedback effects could be important for S0 formation to some extent (in particular for the central star formation of S0s), we will discuss such effects in future work by updating the present code with a self-consistent implementation of AGN feedback.

\subsection{Compact elliptical galaxy}

The elliptical galaxy in our model consists of a dark matter halo and a compact stellar spheroid.  The elliptical is assumed to be initially devoid of gas.  In order to model the dark matter component, we adopt the NFW density profile (\citealt{nfw}) suggested from CDM simulations:
\begin{equation}
{\rho}(r)=\frac{\rho_{0}}{(r/r_{\rm s})(1+r/r_{\rm s})^2},
\end{equation}
where  $r$, $\rho_{0}$, and $r_{\rm s}$ are
the spherical radius,  the characteristic density of the dark halo, and the scale length of the halo, respectively.  We choose reasonable values for the virial radius $r_{\rm vir}$ and the concentration parameter $c=r_{\rm vir}/r_{\rm s}$ based on predictions from cosmological simulations for a given dark halo mass $M_{\rm h, E}$ (e.g., \citealt{neto2007}).

The stellar spheroid is constructed to have the Hernquist density profile
and an isothermal velocity dispersion without global rotation. 
The initial size and the total stellar mass of the spheroid are denoted as $R_{\rm E}$ and  $M_{\rm E}$, respectively. By convention, we choose $R_{\rm E}$ to be the stellar half mass radius. In the present work, $R_{\rm E}$ and  $M_{\rm E}$ are considered to be free parameters that influence the formation processes of S0s.  We mainly investigate models with $M_{\rm h, E}=10^{12}
\Msun$, $M_{\rm E}=6 \times 10^{10} \Msun$, and $R_{\rm E}=0.85$ kpc. The scale length of the Hernquist profile is set to be $R_{\rm E}$ for all models.
 
The adopted value of $R_{\rm E}$ is quite small for the corresponding value of $M_{\rm E}$, particularly when comparing to the population of normal ellipticals observed at $z=0$.  For this reason, we consider our simulated elliptical galaxy to be a compact elliptical.  The adopted values of $R_{\rm E}$ and $M_{\rm E}$ are consistent with the size-mass diagrams for high-$z$ compact systems as given by \citet{damjanov2011} and \citet{dullograham2013}.  We note that the size and stellar mass of our adopted model galaxy is intermediate between the values of local compact ellipticals as studied by \citet{norris2014} and \citet{yildirim2017}. In this context, our initial conditions represent a rather massive and compact elliptical in the high redshift universe.

\subsection{Gas-rich satellite galaxy}

The satellite galaxy in our merger simulations is represented by a bulgeless disc galaxy.
The total masses of dark matter halo, stellar disc, and gas disc of the disc galaxy
are denoted as $M_{\rm h}$, $M_{\rm s}$, and $M_{\rm g}$, respectively. 
In order to describe the initial density profile of dark matter halo, we again adopt the NFW profile (\citealt{nfw}). The radial ($R$) and vertical ($Z$) density profiles of the stellar disc are assumed to be proportional to $\exp (-R/R_{0}) $ with scale
length $R_{0} = 0.2R_{\rm s}$  and to ${\rm sech}^2 (Z/Z_{0})$ with scale
length $Z_{0} = 0.04R_{\rm s}$, respectively.  In this notation, $R_{\rm s}$ represents the `size' of the stellar disc and is equal to five times the radial scale length.  Similarly, the gas disc has a size of $R_{\rm g}$, where the radial and vertical scale lengths are set to $0.2R_{\rm g}$ and $0.02R_{\rm g}$, respectively.  The satellite disc galaxy
is assumed to have $R_{\rm g}/R_{\rm s}=1$ in all models.

Rotational velocities in the disc are caused by the combined gravitational potential of disc and dark halo components.  The initial radial and azimuthal velocity dispersions are assigned to the disc component according to the epicyclic theory with Toomre's parameter $Q = 1.5$. The vertical velocity dispersion at a given radius is set to be half the value of the radial velocity dispersion at that point. The gas mass fraction is denoted as $f_{\rm g}$ ($=M_{\rm g}/(M_{\rm s}+M_{\rm g})$) and is considered to be a key parameter that determines the final structure and kinematics of newly developed disc components of S0s.

The gas disc is assumed to have a metallicity of [Fe/H]$=-0.52$ and the dust-to-metal ratio is set to be 0.4.  These choices resemble a disc galaxy having similar chemical enrichment as the LMC (e.g. \citealt{carrera2011}).  No initial radial metallicity gradient is considered in the present study, because we do not expect such a gradient to significantly affect the morphology and kinematics of S0s formed by the present scenario.  The only impact would be a slight change of the metallicity-dependent cooling of the satellite's gas.

In the present work we primarily discuss the fiducial model for which the key model parameters are summarized in Table \ref{tab:par1} for both the elliptical and satellite.

\subsection{Star formation} \label{sec:sfr}

We adopt the `${\rm H_2}$-dependent' star formation recipe of \citetalias{bekki2013} in which the star formation rate (SFR) is determined by the local molecular fraction ($f_{\rm H_2}$) of each gas particle.  A gas particle {\it can be} converted into a new star if the following three conditions are met: (i) the local dynamical time scale is shorter than the sound crossing time scale (mimicking the Jeans instability) , (ii) the local velocity field is identified as being consistent with gravitational collapse (i.e., $\nabla \cdot {\bf v}<0$), and (iii) the local density exceeds a threshold density for star formation ($\rho_{\rm th}$).
We also adopt
the  Kennicutt-Schmidt law, which is described as 
$ {\rm SFR} \propto \rho_{\rm g}^{\alpha_{\rm sf}}$;  (\citealt{kennicutt98}),
where $\alpha_{\rm sf}$ is the power-law slope.
A reasonable value of
$\alpha_{\rm sf}=1.5$ is adopted for all models.
The threshold gas density for star formation ($\rho_{\rm th}$) is 
set to be 1 cm$^{-3}$ for the fiducial model of the present study.

Each supernova (SN) is assumed to eject a total feedback energy ($E_{\rm sn}$) of $10^{51}$ erg.  Of this total, 90\% and 10\% of $E_{\rm sn}$ are deposited as an increase of thermal energy (`thermal feedback') and random motion (`kinetic feedback'), respectively. The thermal energy is used for the `adiabatic expansion phase', where each SN can remain adiabatic for a timescale of $t_{\rm adi}$. This timescale is set to be $10^6$ yr. We adopt a fixed canonical stellar initial mass function (IMF) proposed by \citet{kroupa2001}, which has three different slopes at different mass ranges. Therefore, chemical evolution, SN feedback, and dust formation and evolution are all determined by the fixed IMF.  Even though high-z compact objects may not have a Kroupa IMF, we do not explore this possibility because we expect IMF variations to have an insubstantial effect on the main results of this paper.

\subsection{Dust and metals}

Chemical enrichment through star formation and metal ejection from SNIa, SNII, and AGB stars is self-consistently included in the chemodynamical code. The code explicitly evolves 11 chemical elements (H, He, C, N, O, Fe, Mg, Ca, Si, S, and Ba) in order to predict both chemical abundances and dust properties. There is a time delay between the epoch of star formation
and those  of supernova explosions and the commencement of AGB phases (i.e., non-instantaneous recycling of chemical elements). We adopt the nucleosynthesis yields of SNe II and Ia from 
\citet{tsujimoto95} and AGB stars from \citet{vdh97} in order to estimate chemical yields.  Despite the fact that the adopted code contains these features, we do not investigate these properties in the present study.

The dust model is  the same as the one adopted in \citetalias{bekki2013} and \citetalias{bekki2014}, described briefly as follows. The  total mass of $j$th component ($j$=C, O, Mg, Si, S, Ca, and Fe) of dust from $k$th type of star ($k$ = SNe Ia, SNe II, and AGB stars) is derived based on the methods described in \citetalias{bekki2013}. Dust can grow through accretion of existing metals onto dust grains with a timescale of $\tau_{g}$. Dust grains can be destroyed though supernova blast waves in the ISM of galaxies and the destruction process is parameterized by the destruction time scale ($\tau_{\rm d}$). We consider models with $\tau_{\rm g}=0.25$ Gyr and $\tau_{\rm d}=0.5$ Gyr.  The reason for this selection is discussed in \citetalias{bekki2013}.

\subsection{${\rm H_2}$ formation and dissociation}

Here we briefly mention the treatment of ${\rm H_2}$ formation and dissociation in the adopted code.  Full details on the formation of ${\rm H_2}$ on dust grains are presented in \citetalias{bekki2014}. The present chemodynamical simulations include both  ${\rm H_2}$ formation on dust grains and ${\rm H_2}$ dissociation by FUV radiation self-consistently. The temperature ($T_{\rm g}$), hydrogen density ($\rho_{\rm H}$),  dust-to-gas ratio ($D$) of a gas particle, and the strength of the FUV radiation field ($\chi$) around the gas particle are calculated at each time step so that the fraction of molecular hydrogen ($f_{\rm H_2}$) for the gas particle can be derived based on the ${\rm H_2}$ formation/destruction equilibrium conditions. The SEDs of stellar particles around each $i$-th gas particle  (thus ISRF) are first estimated from ages and metallicities of the stars by using stellar population synthesis codes for a given IMF (e.g., \citealt{bruzual2003}). Then the strength of the FUV-part of the ISRF is estimated from the SEDs so that $\chi_i$ can be derived for the $i$-th gas particle. Based on $\chi_i$, $D_i$, and  $\rho_{\rm H, \it i}$ of the gas particle, we can derive $f_{\rm H_2, \it i}$ (see Figure 1 in \citetalias{bekki2013}). Thus each gas particle has $f_{\rm H_2, \it i}$, metallicity ([Fe/H]), and gas density.  The total dust, metal, and ${\rm H_2}$ masses are estimated from these properties.

\subsection{Galaxy merging}

The orbit of the satellite galaxy is set to be initially in the $xy$ plane of the elliptical in all simulations. The free parameters determining the orbit include the initial distance between the centre of masses of the satellite and elliptical ($R_{\rm i}$), the pericenter ($R_{\rm p}$), and the orbital eccentricity ($e_{\rm o}$).  The spin of the satellite galaxy is specified by two angles $\theta$ and $\phi$, where $\theta$ is the angle between the $z$ axis and the angular momentum vector of the disc, and $\phi$ is the azimuthal angle measured from $x$ axis to the projection of the spin angular momentum vector onto the $xy$ plane. 

We mainly investigate the models with the following configuration: $\theta = 150\deg$, $\phi = 45\deg$, $R_{\rm i}=52.5$ kpc, $R_{\rm p}=17.5$ kpc, and $e=0.7$. Although we have investigated numerous models with different mass-ratios ($m_2$) of the two galaxies and different orbital configurations, we show only the representative models for which model parameters
are briefly summarized in Tables \ref{tab:par1} and \ref{tab:par2}.

\subsection{Limitations}
The adopted assumption of no gas in the central elliptical is not entirely realistic in comparison to observed systems.  The hot gas in the halo of the compact elliptical would be able to strip cold gas from the infalling gas-rich disc galaxy to some extent, which is not modeled properly in this study.  However, as shown in B14, such ram pressure stripping of low-mass disc galaxies is very limited in halos with total mass of $\sim 10^{12} \Msun$ owing to relatively weaker ram pressure. Accordingly, we do not expect that the exclusion of hot halo gas around compact elliptical galaxies would dramatically influence the present results.

In addition, the assumption of zero initial global rotation in the compact elliptical is not realistic in comparison to observed systems.  However, the lack of initial rotation would not influence the merging processes of satellite galaxies. Nevertheless, the kinematics of the final bulge component of the simulated S0 would strongly depend on the initial kinematics of the compact elliptical.

\begin{figure}
   \begin{center}
   \includegraphics[width=0.5\textwidth]{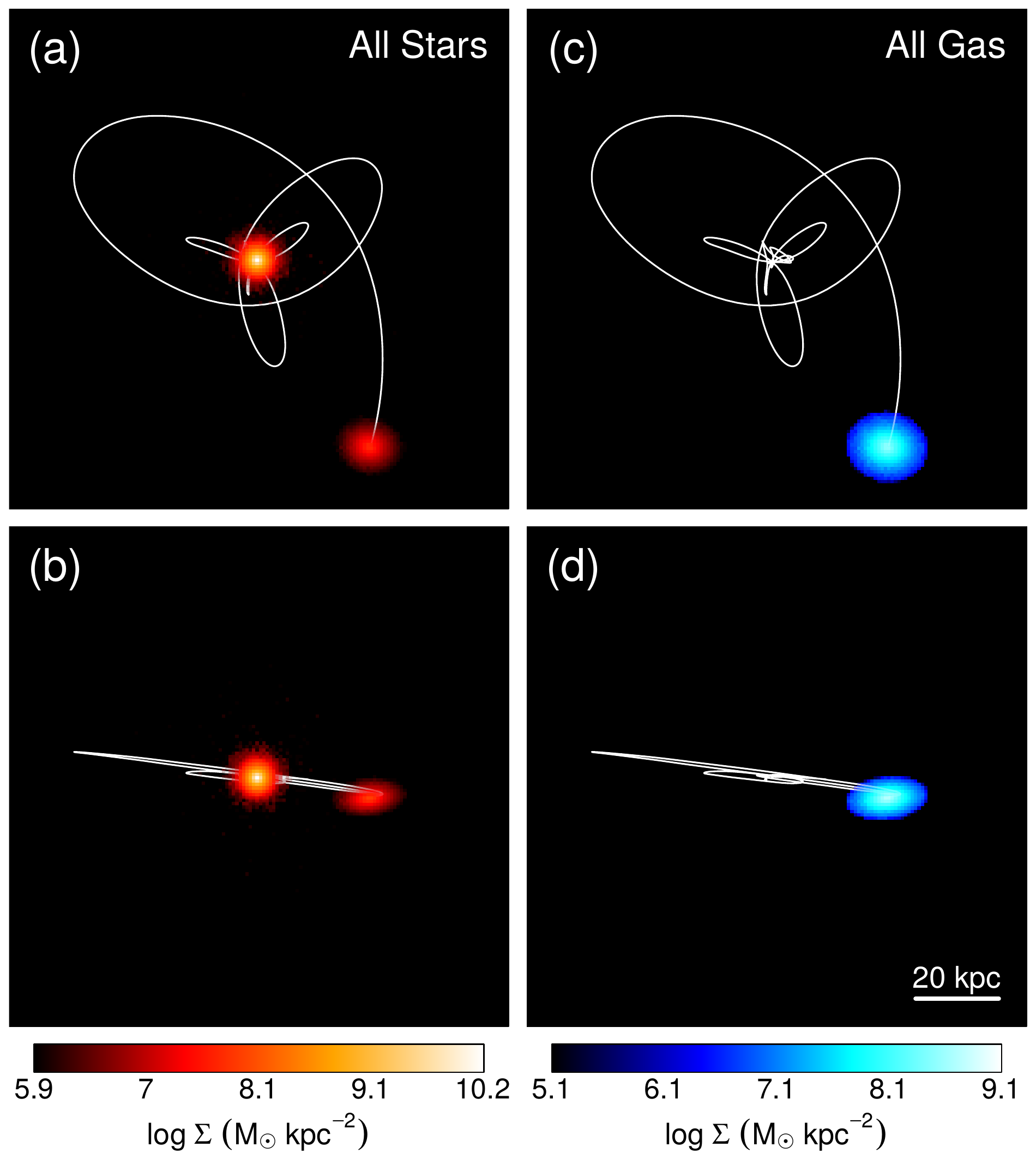}
   \caption{The initial configuration of the fiducial model (at time $t=-2.82$ Gyr) along with the orbit traced by the satellite (white lines) throughout the simulation.  Surface densities of the stellar and gaseous components are shown on logarithmic scales: (a) all stars in a face-on projection, (b) all stars in an edge-on projection, (c) all gas in the face-on projection, and (d) all gas in the edge-on projection.  The face-on and edge-on planes are determined by the total baryonic angular momentum of the disc at the final time step. See the online supplementary material for an animation of these panels from $t=-2.82$ Gyr to $t=0$.}
   \label{fig:simini}
   \end{center}
\end{figure}

\begin{figure}
   \begin{center}
   \includegraphics[width=0.5\textwidth]{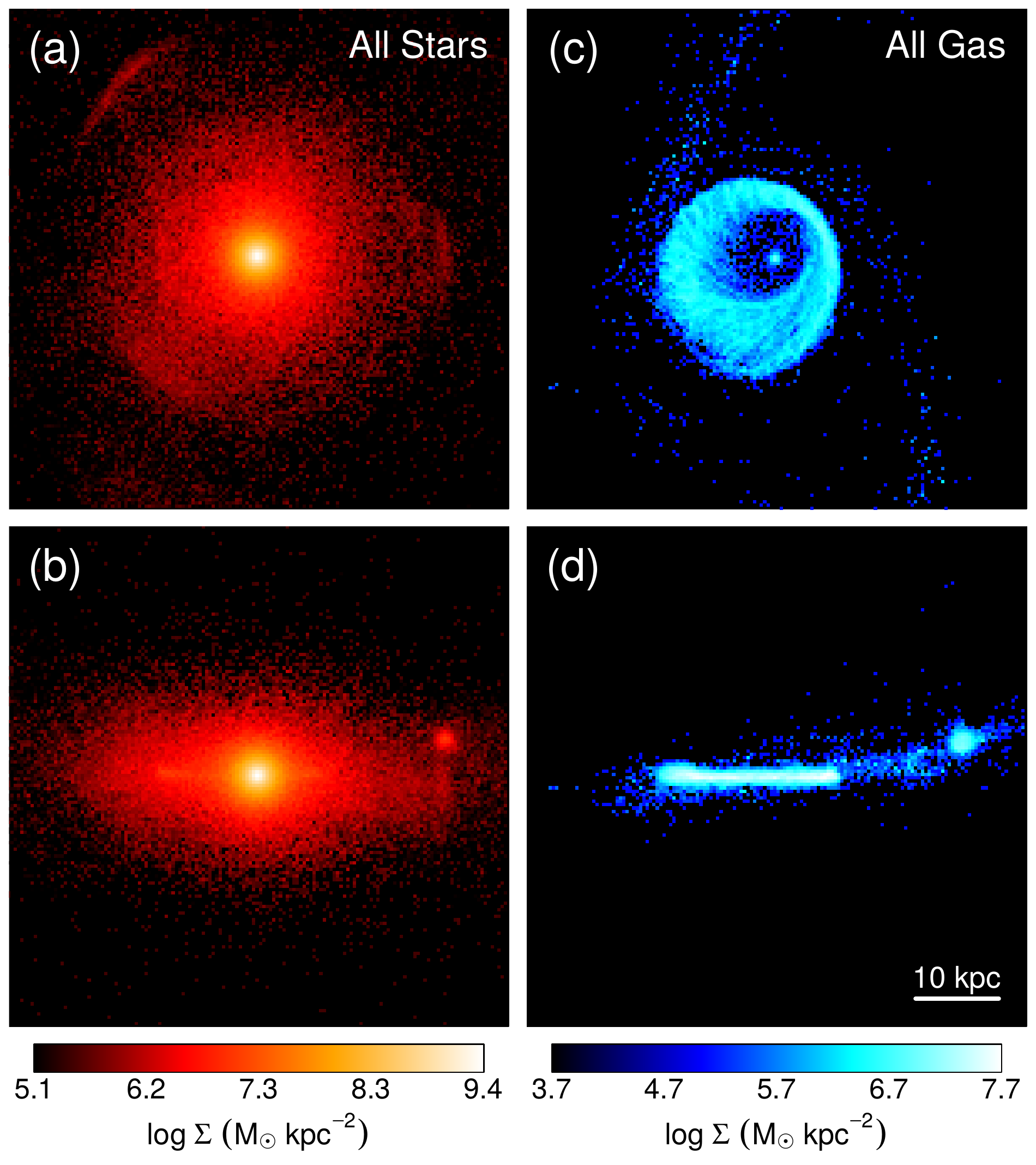}
   \caption{The final configuration of the fiducial model (at time $t=0$ Gyr).  Surface densities are shown for (a) all stars in the face-on projection, (b) all stars edge-on, (c) all gas face-on, and (d) all gas edge-on.  Note that the planes used for each projection are the same as those in Figure \ref{fig:simini} except that the physical scale of each panel is smaller by a factor of 2 as indicated by the scale marker. Salient features of the model include the formation of a disc around the elliptical, a hole within the gaseous disc, and a baryon-dominated dwarf expunged from the satellite by tidal forces (evident in panels (b) and (d) but beyond the range of panels (a) and (c).)}
   \label{fig:simfin}
   \end{center}
\end{figure}

\begin{figure*}
   \begin{center}
   \includegraphics[width=0.8\textwidth]{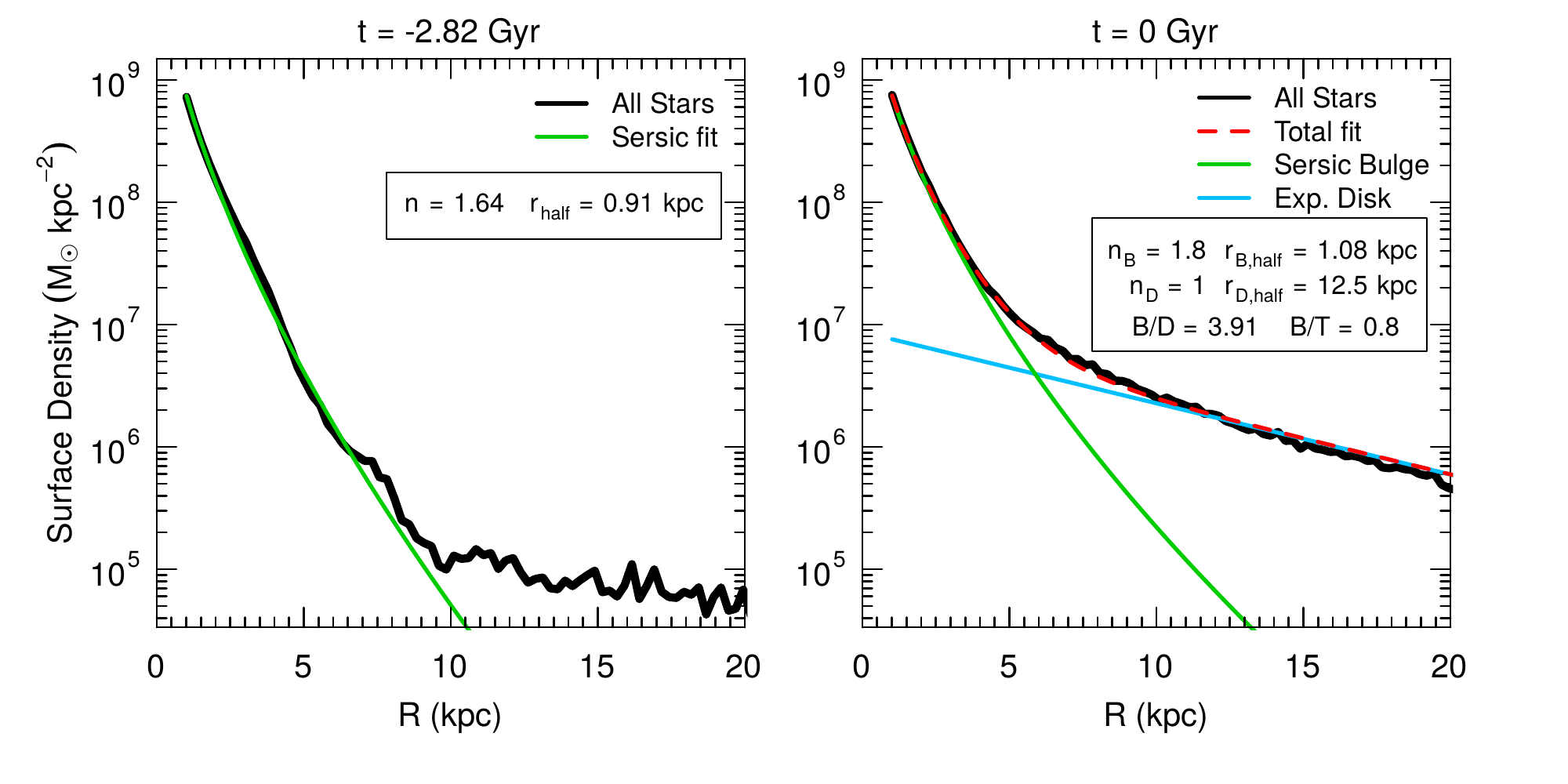}
   \caption{(\textit{Left panel}) One-dimensional stellar surface density of the fiducial model at the start of the simulation ($t=-2.82$ Gyr) (black line) as measured from the center of the elliptical galaxy.  The fit is given by a single Sersic component with free index $n$ (green line), and the boxed inset lists the important parameter values of the fit. (\textit{Right panel}) One-dimensional stellar surface density at the end of the simulation ($t=0$ Gyr) along the cylindrical radius of the disc (black line).  The overall fit (red dashed line) is given by the sum of a bulge (Sersic component with free index $\nB$; green line) and exponential disc (Sersic component with index $\nD=1$).  The boxed inset gives the important parameter values of the fit along with the derived mass ratios between the bulge and disc components. See text for details of the fit.}
   \label{fig:sersic}
   \end{center}
\end{figure*}

\begin{figure*}
   \begin{center}
   \includegraphics[width=1.0\textwidth]{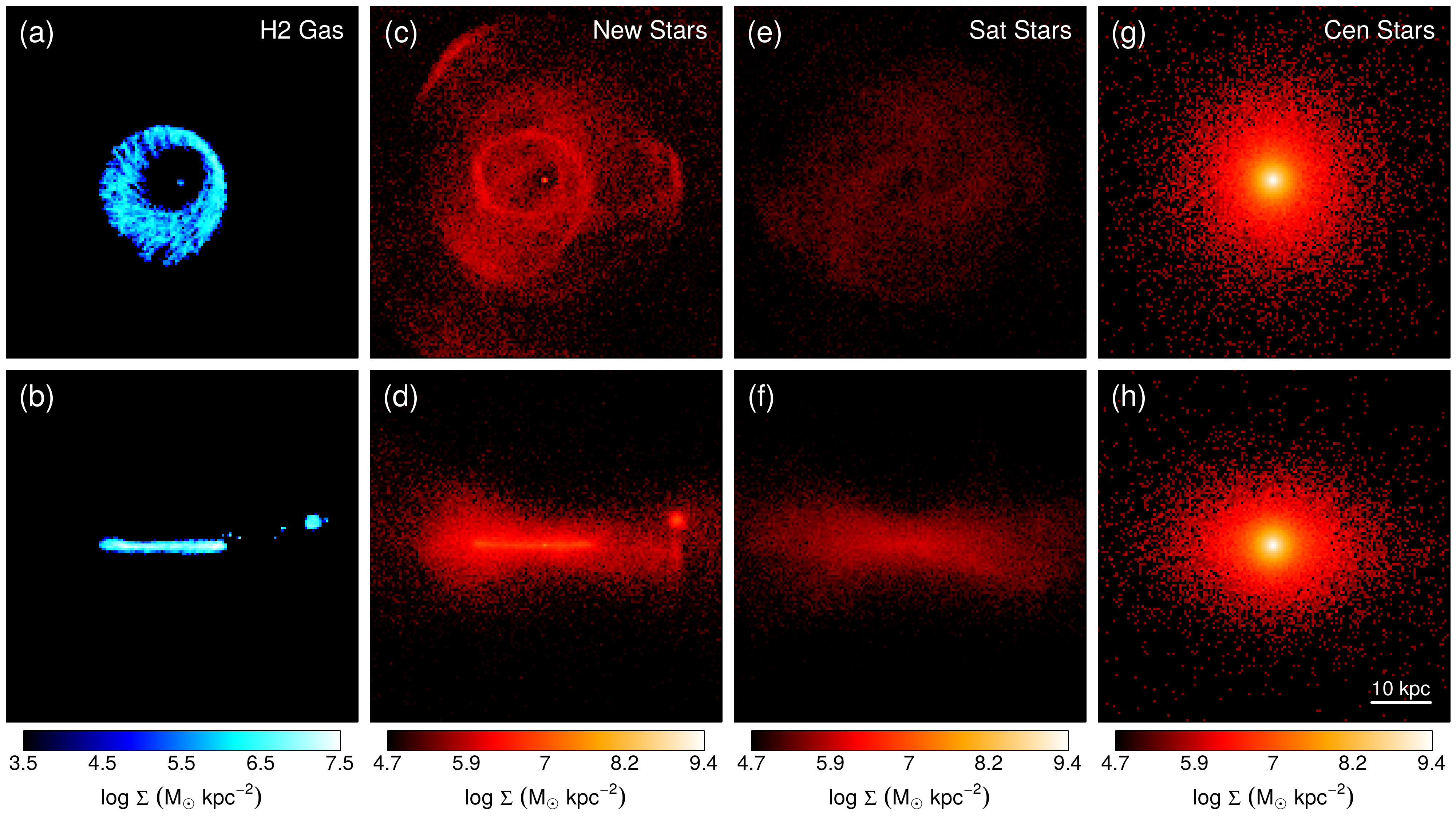}
   \caption{Surface densities of various particles types at the end of the simulation ($t=0$ Gyr): molecular hydrogen shown (a) face-on, and (b) edge on; new stars formed out of gas shown (c) face-on, and (d) edge-on; stars initialized to the satellite shown (e) face-on, and (f) edge-on; and stars of the central elliptical galaxy shown (g) face-on, and (h) edge-on. The planar projections and physical scale are the same as those in Figure \ref{fig:simfin}.}
   \label{fig:ptype}
   \end{center}
\end{figure*}

\begin{figure*}
   \begin{center}
   \includegraphics[width=1.0\textwidth]{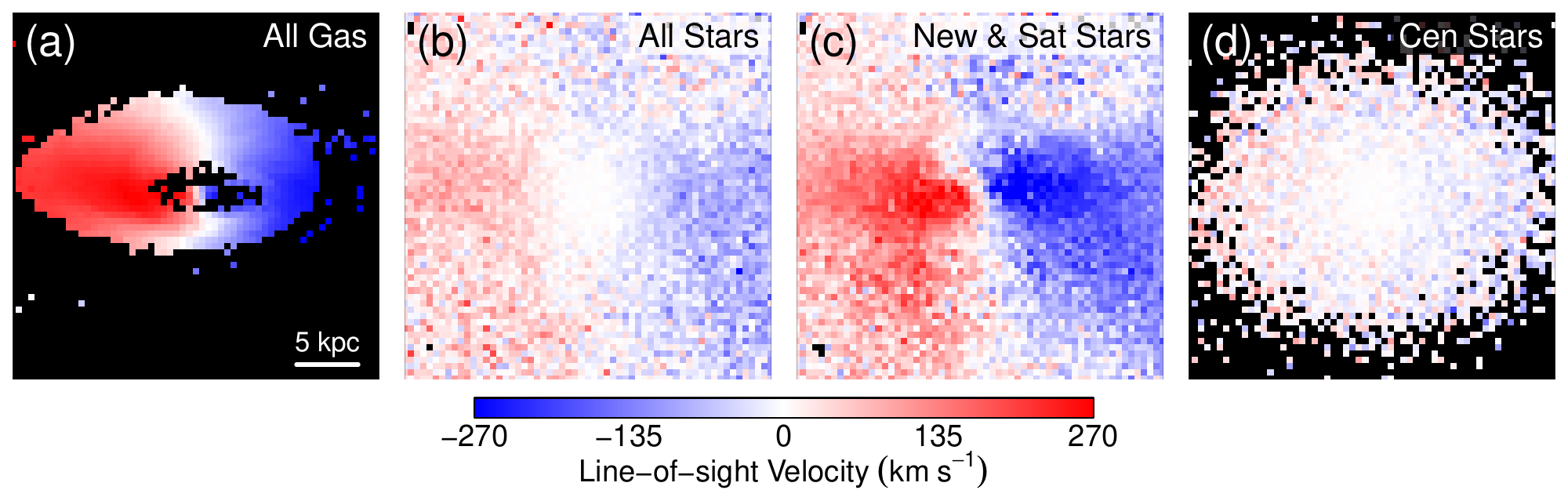}
   \caption{Two-dimensional velocity field of the fiducial model at the final time step with disc inclination set to $60\deg$ with respect to the plane of the sky. The line-of-sight velocity fields are separated by particle type: (a) all gas, (b) all stars, (c) stars formed out of gas plus stars initialized to the satellite, and (d) stars from the central elliptical galaxy. Note that panel (b) is the mass-weighted superposition of the stars in panels (c) and (d). }
   \label{fig:vfield}
   \end{center}
\end{figure*}

\begin{figure*}
   \begin{center}
   \includegraphics[width=0.9\textwidth]{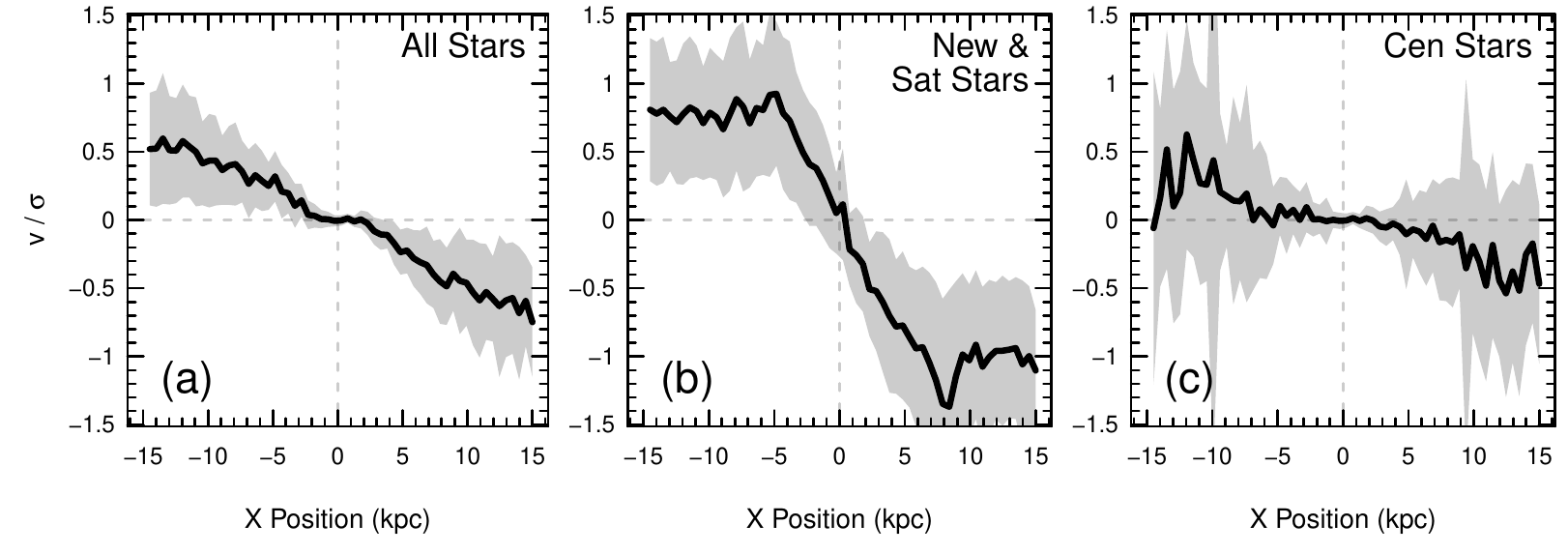}
   \caption{The ratio of velocity to velocity dispersion as a function of x position within the stellar velocity fields of Figure \ref{fig:vfield}. The combined $v/\sigma$ profile for all stars is shown in panel (a), for new and satellite stars in panel (b), and for stars from the central elliptical galaxy in panel (c). The mass-weighted mean value for $v/\sigma$ at each x position is shown as the thick line in each panel, while the grey region denotes the mean plus/minus one standard deviation at each position. Note that the profile in panel (a) is a mass-weighted superposition of the profiles in panels (b) and (c).}
   \label{fig:vsig}
   \end{center}
\end{figure*}

\begin{figure}
   \begin{center}
   \includegraphics[width=0.45\textwidth]{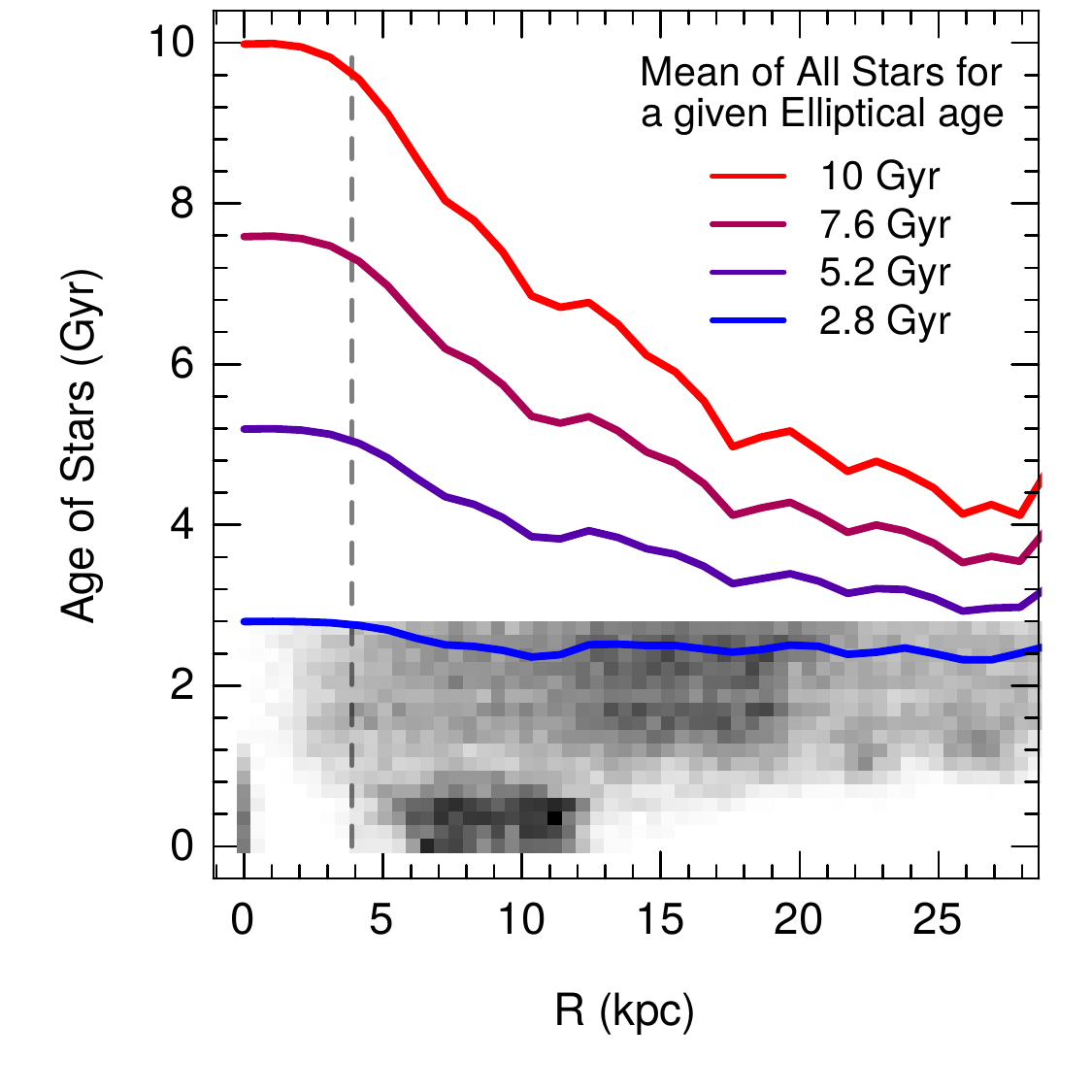}
   \caption{The age of stars as a function of radius in the fiducial model at $t=0$ Gyr. The greyscale plot is a mass-weighted histogram of age versus radius for stars which formed during the simulation, where the greyscale varies from $5 \times 10^4~\Msun$ per bin (white) to $10^7~\Msun$ (black).  These `new stars' have an age which ranges from 0 Gyr (i.e. formed at the end of the simulation) to 2.82 Gyr (i.e. formed at the first time step) as determined by their epoch of formation.  The coloured lines indicate the mass-weighted mean age of all stars in the model as a function of radius.  The lines differ by the assumed age of the stars in the central Elliptical: 2.82 Gyr (blue), 5.21 Gyr (purple), 7.61 Gyr (plum), and 10 Gyr (red).  The vertical dashed line is drawn at twice the stellar half mass radius, indicating the extent of the bulge.  We emphasize for clarity that the greyscale histogram pertains to the newly formed stars only, whereas the coloured lines pertain to \textit{all} stars.  See text for details.}
   \label{fig:age2d}
   \end{center}
\end{figure}

\begin{table}
  \begin{center}
  \caption{Description of the basic parameter values for the simulated compact elliptical and satellite spiral galaxy in the fiducial model.}
  \begin{tabular}{lcc}
  %\hline
  & Elliptical & Satellite \\
  \hline
  
  DM halo mass ($\times 10^{12} \Msun$) & 1.0 & 0.3  \\
  Virial radius (kpc) & 81.7  & 134.2  \\
  $c^\dagger$ &  5 & 12 \\
  Spheroid  mass ($\times 10^{10} \Msun$) & 6.0 & -- \\
  Spheroid  half mass radius (kpc) & 0.85 & --  \\
  Stellar disc  mass ($\times 10^{10} \Msun$) & - & 0.18  \\
  Gas disc  mass ($\times 10^{10} \Msun$) & - & 1.26  \\
  Stellar disc size (kpc) & - & 9.6  \\
  Gas disc size (kpc)  & - & 9.6  \\
  Initial gas metallicity ([Fe/H] dex)  & - & $-0.52$  \\
  Gravitational softening length (pc) & 43 & 138 \\
  Mass resolution ($\times 10^4 \Msun$) & 30.0 & 1.8 \\
  \hline
  \multicolumn{3}{l}{${}^\dagger$ $c$ is the concentration parameter in the NFW dark matter profile.}
  \label{tab:par1}
  \end{tabular}
  \end{center}
\end{table}

\begin{table*}
  \begin{center}
  \begin{minipage}{100mm}
  \caption{Comparison of S0 formation mechanisms in the literature to key observations.}
  \begin{tabular}{lccc}
  \hline
  Mechanism & Environment & {$v/\sigma~$\footnote{$v/\sigma$ is the typical ratio of rotational amplitude to velocity dispersion.  See Section \ref{sec:rkin} } } & {Age gradient\footnote{A positive age gradient implies a young bulge and older disc; a negative age gradient implies an older bulge than disc. Entries ending with a question mark are clarified in the text.  See Sections \ref{sec:rage} and \ref{sec:age}.} } \\
  \hline
  Present work & Field \& low mass groups & $\lesssim 1$ & Negative  \\
  {Ram pressure \& quenching \footnote{\citet{quilis2000,larson80}}} & Clusters \& massive groups & ${\approx 2 - 10}$ & Positive? \\
  {Spiral-Spiral merger \footnote{\citet{tapia2017, borlaff2014, bekki98}}} & Field \& low mass groups & $\approx 1$ & {\brg Positive/Negative?} \\
  {Tidal interaction \footnote{\citet{bekkicouch2011}}} & Groups & $\approx 1$ & Positive \\
  \hline

  \label{tab:comparison}
  \end{tabular}
  \end{minipage}
  \end{center}
\end{table*}

\section{Results: The fiducial model} \label{sec:res}

\subsection{Orbit and merger}

The key parameters of the adopted model are shown in Table \ref{tab:par1} and are discussed in Section \ref{sec:model}.  Figure \ref{fig:simini} shows the initial configuration of the elliptical and satellite along with the orbit traced by the satellite throughout the simulation.  As seen in panels (c) and (d), the elliptical initially has no gas whereas the satellite is gas-rich with a rotation-supported disc. The satellite orbit is mildly eccentric and the orientation of the disc is nearly co-planar with the orbit in a retrograde configuration.
 
As can be seen from the decreasing apocentres in Figure \ref{fig:simini}, the satellite's orbital energy decays throughout the simulation, eventually leading to a merger.  The merger is driven by the dynamical friction encountered by the satellite as it orbits within the dark matter halo of the Elliptical.  While dynamical friction can be modelled analytically (e.g. see \citealt{BT}), the effect is automatically incorporated into our simulation because we self-consistently model all components (both baryon and dark matter) of the satellite and elliptical as `live' N-body systems.

At each successive pericentric passage, stars and gas are removed from the satellite by the strong tidal field of the compact elliptical.  As the merger proceeds, this process of tidal stripping leads to the full destruction of the satellite and the acquisition of its baryons into a disc surrounding the elliptical, as shown in Figure \ref{fig:simfin}.  The final orientation of the disc is determined by the satellite's orbit, such that the orbital plane closely matches the disc plane.  This is simply because the orbital angular momentum is transferred to the angular momentum of the disc.

\subsection{Morphology and structural properties}

The stellar disc in Figure \ref{fig:simfin} is not perfectly smooth but rather exhibits several artifacts.  For instance, there is a shell-like feature in the upper left corner of panel (a) which is formed from the pile-up of stars on similar orbits at apocentre.  Such features are common in tidal debris of destroyed satellites (e.g. tidal streams and shells).  Noticeably lacking in the stellar disc is the presence of spiral arms, which is consistent with S0 morphology.  The lack of spiral structure in our model is by design, as the potential of the central elliptical acts to suppress the disc instabilities which lead to spiral arms (e.g. \citealt{BT}).

The spherical blob evident on the right hand side of panels (b) and (d) of Figure \ref{fig:simfin} is not a feature of the disc but rather lies far behind the disc in projection. It is a baryon-dominated dwarf galaxy that originated as a clump within the satellite and was subsequently removed by the tidal field of the elliptical.  Whereas the satellite merged with the elliptical, the expunged dwarf continues to orbit at large radii within the disc plane.  This behaviour is evident in the online animation which supplements Figure \ref{fig:simini}. For further discussion of this dwarf, see Section \ref{sec:dwarf}. 

As seen in Figure \ref{fig:simfin}, the gas from the satellite settles into a very thin disc around the elliptical.  The stellar disc in comparison has a larger vertical extent and dispersion. This settling of the gas is due to usual gas dynamical effects, namely, the dissipation of energy and angular momentum via fluid viscosity and cooling.

A prominent feature of the gas disc is the `hole' that forms.  It is not a cavity that has been carved out, but rather it is simply due to the fact that the gas flows into the inner region of the galaxy at a specific range of radii { \brf ($5-15$ kpc), as discussed further in \ref{sec:hole}. }

While the formation of a disc in Figure \ref{fig:simfin} may be visually convincing, the surface density must be examined to verify that the stellar distribution is indeed disc-like.  To do so, we first compute the one dimensional stellar surface density profile $\Sigma(R)$ as a function of cylindrical radius $R$ in the disc plane. Then we fit Sersic profiles which take the form
\begin{equation} \label{eq:sersic}
  \Sigma (R) = \Sigma_{e} \exp \left\{ -b_n \left[ \left(\frac{R}{R_e}\right)^{(1/n)}-1 \right] \right\},
\end{equation}
where the free parameters are $n$, the Sersic index, $R_e$, the effective radius, and $\Sigma_{e}$, the surface density at $R_e$ (\citealt{grahamdriver2005}).  The constant $b_n$ can be computed numerically from the equation
\begin{equation}
\Gamma (2n) = 2 \gamma ( b_n, 2n),
\end{equation}
where $\Gamma$ is the (complete) gamma function, and $\gamma$ is the lower incomplete gamma function.

For our purposes, we model a bulge as a Sersic profile $\SigB$ having a free index $\nB$, and we model a disc as a Sersic profile $\SigD$ with index $\nD=1$, which recovers the surface density of an exponential disc.  We fit a single component $\SigB$ to the stellar surface density at the beginning of the simulation ($t=-2.82$ Gyr), and we fit a two-component model $\SigD + \SigB$ to the surface density at the end of the simulation ($t=0$ Gyr).

In this way we can assess the initial properties of the elliptical galaxy and test how those properties change with respect to the formation of a disc by the end of the simulation.  Note that the bulge profile $\SigB$ is fit independently at the beginning and end of the simulation.  We carry out our numerical fit using the AstroPy package with a Levenberg-Marquardt least square fitter (\citealt{astropy}).

Figure \ref{fig:sersic} shows the results of our Sersic fits.  At the start of the simulation, the profile is fit nicely by a single sersic component with $n=1.64$ and $R_e = 0.91$ kpc.  The low density region beyond 10 kpc in the left hand panel of Figure \ref{fig:sersic} is essentially the noise floor from a small number of particles and does not alter the fit.  By the end of the simulation the profile of the bulge changes only slightly, having index $n=1.80$ and effective radius $Re = 1.08$ kpc.  The lack of significant evolution in the bulge profile is confirmation that the compact elliptical is preserved as the core of the final S0. Our value of $n=1.80$ for the Sersic index is somewhat smaller than the typical value of $n=4$ for normal elliptical galaxies (\citealt{dev48}), but it is not unreasonable for a system on the low mass range of elliptical galaxies.  In addition, this value is largely a consequence of our adoption of a Hernquist density profile (see Section \ref{sec:model}).

As seen in the right hand panel of Figure \ref{fig:sersic}, the outer regions of the S0 are nicely fit by an exponential disc profile, and the total two-component fit (red dashed line) matches the surface density very well.  This is confirmation that the satellite merger in our simulation does indeed result in a two-component bulge-disc system.  The total mass ratio between these components favours the bulge, with a value of ${\rm B/D} = 3.91$.  The equivalent bulge-to-total value is ${\rm B/T}=0.80$ which is larger than the average value for observed S0s ($\sim0.6$) but is not atypical for S0s with massive bulges (\citealt{kormendybender2012}).

Figure \ref{fig:ptype} displays the surface densities of our model pertaining to specific particle types.  As detailed in Section \ref{sec:model}, there are three types of stars in our simulation: those assigned to the central elliptical at the start of the simulation, those assigned to the satellite at the start of the simulation, and those which form throughout the simulation from $\H2$ gas.  As seen in panels (a) and (b) of Figure \ref{fig:ptype}, the morphology of the molecular $\H2$ gas is broadly similar to that of the total gas distribution, featuring a central `hole' and the baryon-dominated dwarf.  Panels (c) and (d) indicate that the `new stars' have quite a different distribution to the $\H2$ gas.  This is particularly evident in panel (c) which shows a somewhat patchy distribution of stars highlighted by tidal substructure such as streams and shells.  This contrasts to the smooth distribution of stars that were originally in the satellite disc, shown in panels (e) and (f).  The substructure within the stellar disc is largely hidden by the overlap of other stellar components, particularly at small radii where the elliptical dominates, as shown in panels (g) and (h).

It is evident from inspecting Figure \ref{fig:ptype} that the infalling gas does not significantly fuel the build-up of the bulge.  The highly flattened distributions of $\H2$ gas in panel (b) and of new stars in panel (d) are consistent with the build-up of the disc only. The central elliptical galaxy evolves only marginally by the end of the simulation, acquiring a mild flattening of its outer layers in the equatorial plane as shown in panel (h) of Figure \ref{fig:ptype}.

\subsection{Kinematic properties} \label{sec:rkin}

To indicate the kinematics of the model, we show in Figure \ref{fig:vfield} the two-dimensional velocity field as projected at an inclination of $60\deg$ to the sky.  The gas clearly exhibits disc kinematics with a large amplitude of rotation in panel (a).  The stars also exhibit rotation but with lower amplitude, and primarily in the outer regions as shown in panel (b).  The inner region of the stellar distribution does not exhibit a clear rotational signal because that is where the bulge dominates, rather than the disc.  This can be verified by inspecting the separate velocity fields of each stellar particle type.  Panel (c) indicates a large amplitude of rotation even at small radii for the stars which originated from the satellite and the stars which formed out of the gas.  In contrast, panel (d), shows that the inner parts of the central elliptical exhibit negligible rotation.

Figure \ref{fig:vsig} shows the $v/\sigma$ profiles across the x direction of the stellar velocity fields of Figure \ref{fig:vfield}.  Panel (a) shows $v/\sigma$ for all stars, while panels (b) and (c) show the contributions from the stars that constitute the disc and bulge, respectively. The relative importance of rotation versus random motion is evident in each of these panels.  For instance, the motion of stars within the central elliptical is dispersion dominated, resulting in $v/\sigma \approx 0$ for the inner regions of panels (a) and (c).  As noted previously, the bulge acquires a slight rotation in its outer parts, which is apparent in panel (c) for radii beyond 5 kpc.

Meanwhile, the stars from the satellite exhibit a stronger rotational signal within the disc, with a value of $v/\sigma \approx 1$ beyond 5 kpc in panel (b). This value indicates that rotation and dispersion may be comparably important for the motion of disc stars within our S0 model.

{ \brf Table \ref{tab:comparison} compares the present scenario to other S0 mechanisms regarding $v/\sigma$ predictions.  Our S0 disc has a low $v/\sigma$ value, but such a low value is also expected for the tidal interaction and major merger scenarios.  This kinematic similarity is linked to the gravitational encounters in the field which cause a similar dynamical heating of the disc.  In contrast, an S0 formed from the quenching and/or gaseous stripping of a spiral galaxy in a cluster will retain a high $v/\sigma$ reminiscent of cold discs, as indicated in Table \ref{tab:comparison}. Accordingly, observed values of $v/\sigma$ may be able to discriminate to some degree between different formation channels for S0s, but may be limited by similar predictions for competing mechanisms in the field.}

\subsection{Age profiles} \label{sec:rage}

The distribution of stellar ages within the disc at the end of the simulation is shown in Figure \ref{fig:age2d}.  The greyscale histogram shows the age distribution of `new stars' which form out of the gas during the simulation, from 0 Gyr (i.e. formed at the end of the simulation) to 2.82 Gyr (i.e. formed at the beginning).

While stars of many ages are found in the disc, a dominant portion of the young stars (age $<$ 1 Gyr) are found at radii between $5-12$ kpc in Figure \ref{fig:age2d}.  { \brf This distribution reflects the epoch of recent disc building that was fueled by the merger.  In contrast, the bulge remains largely unchanged for many Gyr and is dominated by older stars.

In contrast to the stars which form out of gas throughout the simulation, the stars which constitute the initial equilibrium models (i.e. the exponential disk and Hernquist sphere per Section \ref{sec:model}) do not have a formation time.  In other words, the simulation does not constrain the ages of these particle types.  For the purpose of constructing age profiles, we are free to assign the age of each of these equilibrium systems as long as they are not less than 2.82 Gyr (i.e. so their 'formation' preceded the first time step of the simulation).

It makes physical sense for the gas-rich satellite to have similar ages between its gaseous disc and thin stellar disc.  This motivates us to assign an age of 2.82 Gyr to the stars which comprise the initial disc of the satellite.

For the stars belonging to the elliptical, we consider several possible ages from 2.82 Gyr to 10 Gyr, each corresponding to a different physical scenario.  For instance, an age of 10 Gyr corresponds to a compact Elliptical which formed at redshift $z=2$ (e.g. a red nugget) which then merges with a young satellite within the past 1 Gyr.  The corresponding age profile of the $z=0$ remnant is shown as the red curve in Figure \ref{fig:age2d}.  The age profiles corresponding to other choices for the age of the elliptical are also shown: 2.82 Gyr (blue), 5.21 Gyr (purple), 7.61 Gyr (plum), and 10 Gyr (red). }

Assigning a progressively younger age to the elliptical flattens the age gradient as seen in Figure \ref{fig:age2d}.  An age of 2.82 Gyr (blue line) corresponds to a scenario where the elliptical and the merging satellite have essentially the same age.  This could apply, for instance, to a high-$z$ red nugget which merges with a gas-rich satellite at $z=1$.

The age difference between the S0 bulge and disc essentially depends on the epoch of satellite accretion.  If it occurred at high redshift, then the age gradient may be flat or may decline only slightly (blue line in Figure \ref{fig:age2d}).  In contrast, recent satellite accretion onto an ancient host elliptical would produce a more dramatic and more easily detectable age gradient (red line in Figure \ref{fig:age2d}). { \brf The robust prediction of our model is that the age gradient will be negative or zero for a variety of merger scenarios. }

{ \brg The age gradients produced by other S0 formation scenarios are noted in Table \ref{tab:comparison}.  Owing to a relative lack of explicit predictions in the literature, we must infer the expected age gradient for several scenarios.  For instance, a spiral which evolves to S0 via the ram pressure / quenching scenario can build a young bulge by retaining gas in its central region, likely producing a positive age gradient.  Similarly, the merger of two spirals can yield an early-type disk galaxy, but gas will be funneled to the center and build a young pseudo-bulge or bar (e.g. \citealt{athan2016}).  Because the strength of the central starburst can vary with numerous factors (e.g. \citealt{mihos94}), the age gradient could conceivably be either positive or negative for the Spiral-Spiral scenario.}

\section{Discussion} \label{sec:dis}

\subsection{Classifying S0s via age gradients} \label{sec:age}

{ \brf 

Field S0s and cluster S0s appear to have distinct properties regarding the ages of their bulges and discs.  S0s found in clusters tend to have bulges that are younger than their discs (\citealt{bedregal2011, johnston2012}), whereas S0s in the field are observed with older bulges than discs (\citealt{tabor2017, guerou2016}).  In cases where the structural components cannot be separated, the implied age gradients would be positive for cluster S0s, and either flat or negative for field S0s.

The predictions of our model are consistent with negative age gradients: as seen in Figure \ref{fig:age2d}, the trend of ages in the S0 disc can vary from flat to sharply declining depending on the age of the compact elliptical.  The core of the S0 is made up of old stars from the compact elliptical, and the outer regions of the S0 are made up of the acquired disc of gas and young stars. { \brg Similarly, the Spiral-Spiral merger scenario also operates in the field and may produce negative age gradients (Table \ref{tab:comparison}), as long as the merger-induced gas flows do not build a dominant central component of young stars.}

In contrast, a positive age gradient is a natural consequence of many S0 formation mechanisms in dense environments, including ram pressure stripping (\citealt{quilis2000}), truncation of star formation (\citealt{aragon2006}), and tidal fields in clusters and groups (\citealt{byrd90,bekkicouch2011}).  In these cases, a progenitor spiral galaxy loses its gas to a final episode of central star formation, which both starves the disc and builds up the bulge with young stars.

Owing to the differences pointed out above, we postulate that field and cluster S0s are distinct galaxy types which are not formed by the same evolutionary process.  We suggest a simple dichotomy: S0s in clusters are associated with spiral progenitors and evolve via numerous distinct mechanisms in dense environments; whereas S0s in the field are{\brg associated with merger events such as the compact elliptical scenario described in the present work}.  We suggest that these subclasses could be distinguished observationally by the distinct ages of their bulges and discs, or equivalently, by their age gradients as noted above.

}

\subsection{Holes and Rings in the disc} \label{sec:hole}

A salient feature of our model is the large hole present in the gaseous disc as seen in Figures \ref{fig:simfin} and \ref{fig:ptype}, { \brf which is reminiscent of the rings observed in some isolated S0s (\citealt{marino2011, delrio2004}).  It is possible that the ring in our model is representative of observed galaxies which were a part of the red sequence but subsequently moved blueward into the green valley due to a merger-induced phase of disc building (\citealt{thilker2010, bouchard2010}).

While consistent with our merger scenario, the presence of HI and UV rings in observed S0 galaxies can be attributed to other sources including: continuous accretion of intergalactic gas (\citealt{moiseev2010}), bar instabilities which transport gas to the core and outer radii (\citealt{marino2011}), and the gradual fading of original star formation activity (\citealt{bresolin2013}).

The 'hole' in our model is created during the infall of the gas as it settles into the disc plane.  Because the source of the disc gas is a merging satellite on a moderately eccentric orbit, the infalling gas retains is orbital angular momentum and settles into a ring in the disc plane.  Without a mechanism to dissipate this angular momentum, the gas is not capable of flowing into the center of the galaxy and cannot 'fill the hole'.  Similar ring morphologies are evident in previous numerical studies of minor mergers (\citealt{mapelli2015}).

In our model, the central elliptical is idealized in the sense that it does not have a hot gas envelope nor a cold gas disc. However, if the satellite had plunged through a hot envelope or cold disc around the elliptical during merger, then the viscous interactions would dissipate energy and angular momentum and drag the infalling gas toward smaller radii. Such radial inflows would likely fill in any 'holes' that would have otherwise formed.

}

\subsection{Formation of baryon-dominated dwarf} \label{sec:dwarf}

{ \brf

The dwarf seen in panels (b) and (d) of Figure \ref{fig:simfin} has a total mass of $4.12 \times 10^8 \Msun$, a baryon fraction of 99.7\%,  a gas fraction of 64.0\%, and a stellar half mass radius of 1.1 kpc.  These figures imply that the dwarf is too compact to be an ultra diffuse galaxy (\citealt{vandokkum2015}), far too depleted of dark matter to be a low surface brightness dwarf (e.g. as in the Local Group, e.g. \citealt{mcc2012}), and too large to be an Ultra Compact Dwarf (\citealt{norris2014}).  Rather, this dwarf is more akin to a tidal dwarf galaxy (TDG) in both total mass and baryon fraction (e.g. \citealt{duc2004}).

However, the dwarf is distinct from TDGs for a subtle reason.  TDGs form within tidally expelled debris following a galactic interaction (\citealt{dab2013}), but the dwarf in our simulation forms prior to any tidal encounter.  Gas instabilities in the disc of the spiral trigger the coalescence of a clump of baryons, similar to the fragmentation observed in high-redshift discs (e.g. \citealt{bournaud2008}).  But rather than migrating and disrupting, the clump in our simulation is subsequently expelled from the spiral by the tidal field of the elliptical.
 
This `clump dwarf' is also striking because it orbits within the disc plane of the final S0 system.  We speculate that S0s formed by the mechanism described in this paper may have an enhanced likelihood of harbouring a dwarf satellite in the same plane as the S0 disc.  This conjecture may well apply to systems not represented by our simulation, such as surviving group members from the infall and subsequent merger of a co-orbiting group of dwarfs.

}

\subsection{Model improvements}

{ \brf In this work we have presented a simple numerical model to illustrate a new pathway for S0 formation.  Here we consider limitations of our model. }

We have considered the case of a single satellite merger only, but the possibility of multiple mergers should be investigated.  In the context of cosmological mass assembly, multiple mergers can be delivered naturally by group infall onto the elliptical.  The repeated infall of multiple small satellites on roughly similar orbits would contribute to the buildup of the S0 disc by providing a steady source of mass and angular momentum.  It would also reduce the B/T ratio compared to the somewhat large value of 0.8 for our fiducial model (e.g. see Figure \ref{fig:sersic}).

{\brf Because our treatment of the elliptical galaxy is simplistic, future work should explore more realistic treatments, including: } adopting an initial rotation profile, assigning gaseous components (whether cold disc or hot halo), and exploring different radial density profiles.  In particular, the degree to which our S0 formation scenario depends on the elliptical being gasless and non-rotating should be quantified, as it may affect the density profile and age gradient within the disc.

{ \brf Our best model has a satellite-to-elliptical mass ratio of 0.3, but other favourable models could potentially have mass ratios anywhere in the range of $0.2-0.5$ (see Appendix \ref{sec:alt}).  The exact range would likely depend on the effect of varying other parameters such as orbital eccentricity, disc orientation, gas fraction, etc.  Several such alternate models are briefly considered in Appendix \ref{sec:alt}.

It remains the task of future work to fully explore the parameter space of possible interaction histories for our S0 formation scenario.  The question of how often or how likely this transformation occurs in the Universe in comparison to other S0 mechanisms is likewise left for future study.  For instance, one can quantify the frequency of mergers between ellipticals and gas-rich satellites in comparison with other galaxy types by considering observed galaxy number densities. }

\section{Summary and Conclusion} \label{sec:sum}

{ \brf In this paper we have presented a new scenario for creating S0s which we illustrate with a hydrodynamical N-body model.  Whereas S0 formation scenarios in the literature typically assume a spiral progenitor, our scenario starts with a compact elliptical.  Following the merger of a gas-rich satellite, a disc is built up around the compact elliptical, effectively transforming it to an S0.  Notably, this process can occur in low-density environments like the field in contrast to many S0 formation mechanisms in the literature which may require dense environments like clusters. }

In our model, the compact elliptical evolves only marginally to become the central bulge of the S0. Thus we form a classical bulge with properties similar to an elliptical galaxy. The potential of the bulge plays an important role in determining the morphology of the disc.  Namely, the stellar disc has no apparent spiral structure, which is crucial for the final system to be considered S0.  This is simply due to the fact that disc instabilities are suppressed by the potential of the bulge, and therefore spiral structure cannot be maintained.

{ \brf We predict } that the age of stars should exhibit a clear gradient in the S0.  The bulge should be the oldest structural component and the disc should be the youngest, leading to a negative age gradient as a function of radius in the S0.  The steepness of this gradient depends on the age of the elliptical as shown in Figure \ref{fig:age2d}.  {\brg The negative age gradient may be a distinguishing characteristic of the present scenario, but further work needs to be done to clarify the predicted age gradients of other S0 formation mechanisms in the literature (e.g. Table \ref{tab:comparison}).}

A key conclusion to be drawn from this work is that S0s may be a catch-all category of rather diverse systems.  We postulate that S0s may be separated into two subclasses by their distinct evolutionary paths: those which evolved from spiral galaxies in dense environments, and those which{\brg formed via mergers in sparse environments, including the compact elliptical progenitors of the present scenario. }  As discussed in Section \ref{sec:age}, observed age gradients may be a key discriminator for disentangling these subclasses.

\section*{Acknowledgements}

{ \brf We thank the anonymous referee for constructive feedback which led to the improvement of our paper.} This research was supported by the Australian Government through the Australian Research Council's Discovery Projects funding scheme (DP170102344).

\bibliographystyle{mnras}
\bibliography{s01}

\appendix
\section{Alternate Models} \label{sec:alt}

\begin{table}
  \begin{center}
  \begin{minipage}{80mm}
  \caption{Description of the parameter values for alternate models M2-M6 in comparison to the fiducial model M1.}
  \begin{tabular}{lccc}
  \hline
  Model ID & {$m_2$\footnote{$m_2$ is the dynamical mass ratio between the satellite and elliptical.}} & {$\rho_{\rm th}$ (cm$^{-3}$)\footnote{$\rho_{\rm th}$ is the star formation threshold gas density as in Section \ref{sec:sfr}.} } & {$\theta$ ($\deg$)\footnote{$\theta$ is the angle between the orbital and spin angular momentum vectors. $\theta=0\deg$ is the prograde case, and $\theta=180\deg$ is retrograde.} } \\
  \hline
  {M1\footnote{Model M1 is the fiducial model.} } & 0.3 & 1 & $150$ \\
  M2 & 1.0 & 1 & $150$ \\
  M3 & 0.5 & 1 & $150$ \\
  M4 & 0.1 & 1 & $150$ \\
  M5 & 0.3 & 10 & $150$ \\
  M6 & 0.3 & 1 & $0$ \\
  \hline

  \label{tab:par2}
  \end{tabular}
  \end{minipage}
  \end{center}
\end{table}

\begin{figure}
   \begin{center}
   \includegraphics[width=0.45\textwidth]{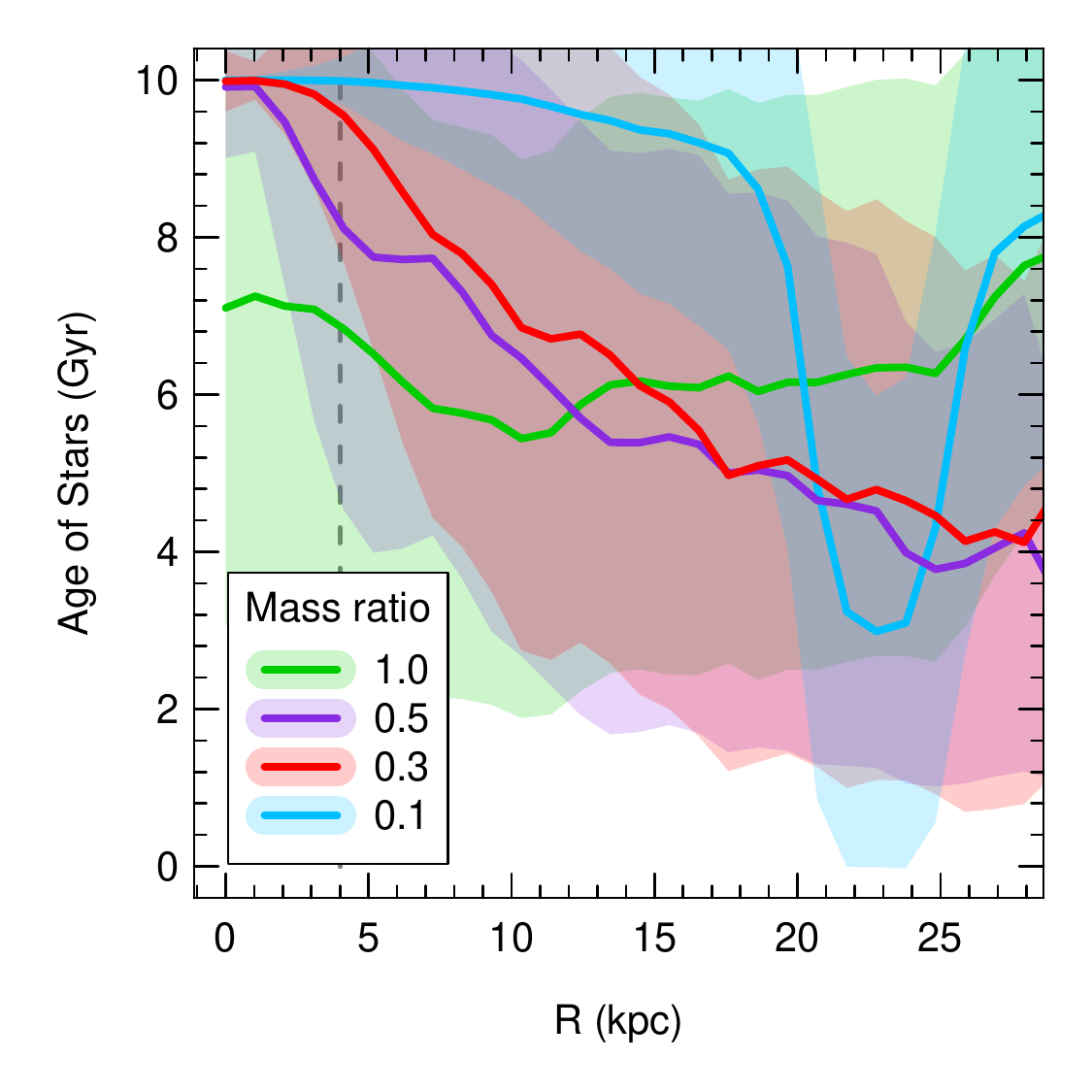}
   \caption{Stellar ages as a function of radius in the disc for four different models.  Each model has a different mass ratio between the satellite and central elliptical galaxy: a mass ratio of 1.0 (green), 0.5 (purple), 0.3 (the fiducial model; red), and 0.1 (blue). In Table \ref{tab:par2}, these models are labeled as M2, M3, M1, and M4, respectively.  Solid lines indicate the mass-weighted mean stellar age, and the shaded regions indicate the statistical spread in ages as the mean plus/minus one standard deviation.  The age of stars belonging to the compact Elliptical in each case is assumed to be 10 Gyr.  The vertical dashed line is drawn at twice the average stellar half mass radius, indicating the average extent of the bulge in the models.  The red line in this figure is identical to the red line of Figure \ref{fig:age2d}.}
   \label{fig:age}
   \end{center}
\end{figure}

\begin{figure}
   \begin{center}
   \includegraphics[width=0.5\textwidth]{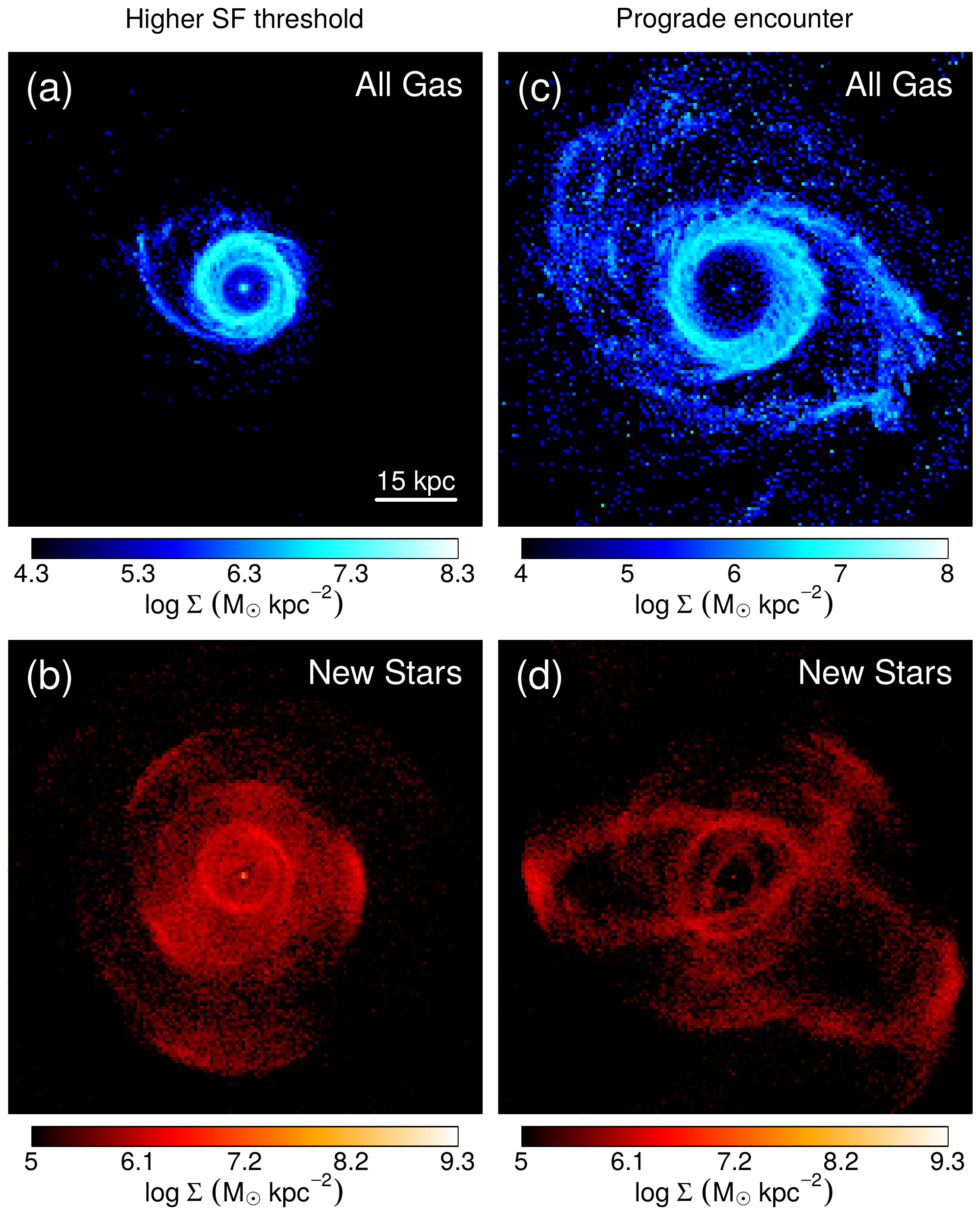}
   \caption{Face-on surface densities of gas and stars in two alternate models.  The consequences of increasing the star formation threshold of the fiducial model is shown for (a) all gas, and (b) stars formed out of the gas.  The consequences of changing the orientation of the satellite disc from a retrograde to a prograde encounter is shown for (c) all gas, and (d) stars formed out of the gas.  In each panel, the face-on projection is determined by the orientation of the final disc.  In the parameter list of Table \ref{tab:par2}, these two alternate models are labeled M5 and M6 for the model with higher star formation threshold and for the prograde model, respectively.}
   \label{fig:other}
   \end{center}
\end{figure}

{ \brf

For the S0 formation mechanism presented in this paper, a vast range of configurations and parameterizations can be investigated for the gas-rich satellite and the massive compact elliptical.  For instance, we can consider variations over the masses, rotational properties, baryon fractions, gas phases, orbital parameters, etc.  Some of these interactions will trigger a transformation into an S0 while others will not.  Although an exhaustive parameter study is beyond the scope of the current work, we briefly discuss the features of a few alternate models to consider the sensitivity of the adopted scenario to different physical conditions.

\subsection{Mass ratio}

The mass ratio between the satellite and elliptical galaxy controls the merger timescale as well as the amount of mass which is available to build a disc.

Figure \ref{fig:age} shows age as a function of disc radius for various models with different mass ratios between the satellite and elliptical.  The red curve represents the fiducial model (mass ratio of 0.3) and is the same as the red line in Figure \ref{fig:age2d}.  In choosing the best model, we seek S0s which have old bulges and young discs as observed in the field (e.g. \citealt{tabor2017}) which translate to a negative age gradient in Figure \ref{fig:age}.  Judging by this criterion, a mass ratio of 0.5 (purple line) is the only other acceptable case along with the fiducial model (red line). 

Given that our S0 formation mechanism relies on a satellite merger, smaller satellites are not ideal candidates since they have less mass for building the S0 disc.  In addition, because of the reduced impact of dynamical friction, the mass ratio of 0.1 (blue line) has not yet produced a merger, indicated by the large spike between $20-25$ kpc in the blue curve of Figure \ref{fig:age}.  Following the orbital evolution for $6-8$ Gyr would produce a merger (\citealt{mbk2008}), but the remnant would not transform to an S0 owing to the relative lack of material for building a disc around the elliptical.

In the case of more massive satellites, the orbits decay too quickly and the delivery of gas to the central elliptical is too rapid to coalesce into a disc.  The material is instead funneled to the center of the galaxy with little overall angular momentum, where it fails to contribute to the buildup of a disc.  This effect is reflected in the equal mass merger case in Figure \ref{fig:age} (green curve).  The buildup of young stars in the inner galaxy causes the central region to be much younger than for other mass ratios.  In addition, the outer regions for the green curve are relatively older in comparison to other mass ratios, which reinforces the notion that the gas and young stars were not able to settle at large radii.

In summary, the mass ratio between the satellite and elliptical is a key parameter in our scenario.  We find it must be large enough ($>0.1$) to lead to a massive merger capable of building a disc.  But it must be small enough ($<0.5$) so that the accreted material does not funnel to the center of the galaxy but rather settles at radii typical of discs.

\subsection{Other parameters}

Figure \ref{fig:other} displays the surface densities of gas and new stars in two alternate models.  In one model, the star formation threshold was raised, and in the other the disc orientation was changed so as to produce a prograde tidal encounter.  Comparing to the surface densities in Figures \ref{fig:simfin} and \ref{fig:ptype}, it is clear that a retrograde encounter and low star formation threshold are important factors in the formation of an S0 with a smooth stellar disc.

As seen in panel (a) of Figure \ref{fig:other}, imposing a higher star formation threshold yields a gas distribution which is qualitatively similar (e.g. a ring-like morphology) but is more centrally concentrated than the fiducial model.  Meanwhile, the stellar disc in panel (b) exhibits tidal shells rather than a smooth disc.  The shells are located at a variety of radii and position angles, reminiscent of the dramatic tidal shells of NGC 474 (e.g. \citealt{turnbull99}).  The visual similarity implies that at least some of the shells of NGC 474 were formed in this way, namely, via the gradual tidal dissolution of a single satellite.

Panels (c) and (d) of Figure \ref{fig:other} show that a prograde tidal encounter does not produce an S0 because the stellar morphology is not disc-like.  The stars which form out of the gas are distributed in sharply defined tidal streams, which is a consequence of the well-known fact that tidal stripping is more efficient on prograde orbits (e.g., \citealt{toomre72}) as compared to the retrograde tidal encounter of our fiducial model.

}

\end{document}